\documentclass{elsart}
\usepackage{amssymb}
\usepackage{amsmath}
\usepackage{graphics}

\begin{document}
\begin{frontmatter}

\title{\large \bf A Possibility to Observe Short-Range NN Properties
 in the Deuteron Breakup $pd\to ppn$ }

\author[dubna,almata]{Yu.N.~Uzikov}\footnote{E-mail address:
 uzikov@nusun.jinr.dubna.su}
\address[dubna]{JINR, LNP, Dubna, 141980, Moscow Region, Russia}
\address[almata]{ Kazakh State University, Almaty,  480121 Kazakhstan}

\begin{abstract}
 Quasi-binary   reaction of the deuteron breakup
 $p+d\to (pp)+n$  with 
 the final proton-proton  pair $(pp)$ in the  $^1S_0$
  state is analyzed  at initial energies $0.5 - 2$ GeV
 in the kinematics of backward elastic pd-scattering $pd\to dp$.
 On the basis of the main mechanisms  of the $pd\to dp$ process, including 
 initial and final state interaction,  we show
 that unpolarized cross section and  spin observables of this reaction
 exhibit  important
 properties of the half-off-shell  $pp(^1S_0)$-scattering amplitude, which
 are relevant to the nucleon-nucleon interaction at short distances.

\end{abstract}
\vspace{5mm}
\vspace{8mm}
\noindent
\begin{keyword}
  deuteron breakup, nucleon-nucleon interaction
\begin{PACS}
24.70.+s; 24.50.+g; 21.45.+v \\[1ex]
\end{PACS}
\end{keyword}
\end{frontmatter}

 The short-range structure of the lightest nuclei
 is related to  fundamental problems of the theory of strong interactions.
  At present the main task  consists  in a clear experimental
 observation of this structure  and determination to what  limiting values
 of  the intrinsic nucleon momenta  $q$  the traditional description
 of nuclei in terms of nucleons is valid.
 In the framework of  the impulse approximation (IA),
  available inclusive experimental data on spin averaged cross section
 for   deuteron disintegration
  ${ d}\,p\to p(0^\circ$)\,X, ${ d}\,^{12}$C$\to p(0^\circ$)\,X (see 
\cite{dubnainclus,dubna} and references therein) are
 compatible with  the realistic deuteron wave functions  at  low
 internal momenta  $q<0.3$ GeV/c only. At higher momenta 
 a systematic deviation from the IA is observed \cite{perdrisat90}.
  This disagreement  is very strong  for the
  tensor analyzing power $T_{20}$ of these reactions.
 On the contrary, the recent JLab data on $T_{20}$ in the elastic 
 ed-scattering \cite{abott} demonstrate that the similar
 impulse approximation with standard wave function, such as RSC \cite{rsc},
 Paris \cite{paris} or
 Bonn \cite{bonn}, is in fair agreement  with the experiment over
 the interval of transferred momenta $Q^2=0 - 1.7 $ (GeV/c)$^2$,
 i.e. for the same internal momenta  $q\sim Q/2\sim 0.7$
 GeV/c as probed in pd-interaction.
 On the whole, such difference between  the ed- and pd- interactions
 can be attributed to a presence
 of specific for pd-collisions  mechanisms, in particular, to the exotic
 like $N^*$-exchanges in the $u$-channel \cite{kk,kob98} or
 three-baryon resonances in the $s$-channel \cite{lak}.
  The deviation from the IA in exclusive data on the reaction
 $p+d\to p+n+p$ can be  explained in part by rescatterings
 in the final state \cite{aleshin}. Other
 ordinary mechanisms like double pN-scattering
 with excitation (and de-excitation)  of nucleon isobars ($\Delta, N^*$)
 in the intermediate states can give a large contribution  also.
 The latter mechanisms are less sensitive to the
 short-range structure of the deuteron than the IA mechanism. At present
 a correct treatment of these mechanisms is a nontrivial problem
 for theory, as it can be seen from the analysis of backward elastic pd-
 scattering in GeV region \cite{lak,kolybas,bdillig,uz98}.

 To minimize  screening effects caused  by excitations of nucleons inside
 the deuteron due to its interaction with incident proton we propose
  to study the deuteron breakup
  with formation of the final spin singlet 
 $(NN)_s$ pair in the $^1S_0$ state in kinematics of the backward
 (quasi)elastic pd-scattering. These are 
\begin{equation}
\label{reaction1}
 p+d\to (pp)+n
\end{equation}
 and $n+d\to (nn)+p$  reactions
 at small relative energy
  of the final NN-pair, $E_{NN}=0-3$ MeV \cite{smuz98}.
 According to  \cite {imuz90}, the presence of the  $pn(^1S_0$)-pair
 in the final state  provides a  considerable 
 suppression of the amplitude of the $\Delta$-isobar mechanism
 due to isospin factor of $\frac{1}{3}$ 
 in comparison with the  elastic backward pd-scattering.
  We note here   that the same suppressing isotopic spin
  factor 
 is effective for  broader class 
 of  diagrams than  established in \cite{imuz90}.
 Furthermore, we  take into account initial and final state interactions
 in the eikonal approximation.
 In addition,  we present the results of calculation of some spin
 observables.

 The dynamics of the reaction (\ref{reaction1}) is discussed here
 by analogy with   the backward elastic pd-scattering
\cite{lak,kolybas} taking into account one nucleon exchange (ONE),
 single $pN$ scattering (SS) and double $pN$ scattering with excitation
 of the $\Delta$ isobar in the intermediate state (figure 1 ({\it a-c})).
 As was shown in \cite{uz98,uz2000}, at  low kinetic
 energies of the proton beam $T_p\sim 0.2-0.3$ GeV
 the coherent sum of  ONE+SS describes  energy and angular dependence of the
 cross section  and  explains qualitatively vector and tensor analyzing
 powers
 of the $pd\to dp$ process in backward hemisphere.
 At higher energies $T_p=0.4-1.0$ GeV  the $\Delta$-mechanism dominates.
 A $\pi+\rho$  exchange model
 of the NN$\rightleftharpoons \Delta$N  amplitude \cite{imuz88} describes
 well
 the experimental cross section of
 the pp$\to pn\pi^+$ reaction \cite{habitzsch} in the $\Delta$-region.
 The  ONE+$\Delta$+SS model with the same $\Delta$-amplitude as in
  \cite{imuz88}, allows one to describe the energy dependence
 of the $pd\to dp$ cross section in the interval of energies
 $T_p= 0.5-2.5$ GeV at $\theta_{cm}=180^\circ$.

 A detail formalism for the amplitude of the reaction $pd\to (NN)N$
 in the framework of the ONE+SS+$\Delta$ model can be derived from the 
  $pd\to dp$ formalism of Refs. \cite{lak,uz98}. For this 
 aim one should  make the following substitution into the matrix elements
 \begin{equation}
 \label{sub}
 |\varphi_d>\to \sqrt{m}|\Psi_{ k}^{(-)}>,
 \end{equation}  
 where $|\varphi_d>$ is the deuteron final state in the  $pd\to dp$,
$m$ is the nucleon mass and
 $ |\Psi_{k}^{(-)}>$  is the scattering state of the final
 NN-system  at the relative momentum ${ k}$
 in the $N+d\to(NN)+N$ reaction.
 Since  the S-wave gives the main contribution to the singlet $(NN)_s$
 state at   $E_{NN}< 3$ MeV, 
 one should omit the D-component of the final deuteron state
 $|\varphi_d>$ in the $pd\to dp$  formalism  \cite{lak,uz98}
 when making the substitution (\ref{sub}).
  Thus,  one has to insert  into the  upper vertex of the ONE diagram 
(figure 1 ({\it a}))
  the half-off-shell amplitude of $pp$-scattering in the $^1S_0$ state,
  $t_s(q,k)$.
 This amplitude is
 shown  in figure 2 as a function
 of the off-shell momentum $q$ at different values $E_{pp}=\frac{k^2}{m}$.
 The first node of 
 $t_s(q,k)$ at $q=0$ is caused by the Coulomb repulsion, whereas
 the second node
 at $q\sim 0.4 $ GeV/c arises due to  short-range repulsion
 in the $NN$ potential of strong interaction.
 A similar node, available in the deuteron S-wave function in the momentum
 space $u(q)$  at $q\sim 0.4$ GeV/c,  can be  connected indirectly to the
 null  of  the measured deuteron charge formfactor $G_{C}(Q)$ 
 at the  transferred momentum $Q\sim 4.5$ fm$^{-1}$ \cite{nikhef,abott}.  The 
 node of $u(q)$  was not yet observed directly in any  reactions with
 the deuteron  due to  large  contribution of the deuteron D-state. 
 An important feature of the reaction
 (\ref{reaction1}) is a possibility to display  the node
 of the amplitude $t_s(q,k)$  straightforwardly in the cross section.
 One can  see it from the following formula for the cross section
 of the reaction (\ref{reaction1}) derived for the ONE mechanism
 (figure 1 ({\it a})) 
\begin{equation}
\label{onecs}
\frac{d^5\sigma}{d p_1d \Omega_1 d\Omega_2}=
K\left [ u^2(q)+w^2(q)\right ]
 \,|t_s(q^\prime,k)|^2,
\end{equation}
 where
 $w(q)$ is the D-wave of the deuteron state,
 $K$ is the kinematical factor; the indices  1 and 2 refer to the final
 protons which are  detected in the forward direction. 
 Using relativistic kinematics for the ONE amplitude  \cite{smuz98},
 one can find that the node of $t_s(q^\prime,k)$ at $q^\prime\sim 0.4$
 GeV/c
 corresponds to the  energy  $T_p\sim 0.7$ GeV at
 the neutron scattering angle $\theta_{cm}=180^\circ$.

 The isospin factors for the reaction $p+d\to (pn)_s+p$ are discussed
 recently
 in  \cite{uz2000}. As was shown there, an  additional isospin
 factor of $\frac{1}{3}$ arises for the $2\pi$ exchange
 mechanism of this reaction in comparison with the $pd\to dp$
  process.
 This mechanism includes, in particular, the $\Delta-$ and $N^*-$ 
 excitations
 in the intermediate state. On the contrary, for the ONE mechanism the 
 isospin factor equals 1.

 Rescatterings in the initial and final states are taken into account
 here for the  ONE-mechanism 
 using an eikonal version of
 the distorted wave Born approximation (DWBA).
 This  method was successfully
 applied in  \cite{blu}   to the backward
 elastic p$^3$He-scattering at $T_p\geq 1$ GeV in the framework
 of the  np-pair exchange mechanism. Another application was done to
 the backward elastic $pd$ scattering within the ONE mechanism 
\cite{uzpddp98}.  Due to rescatterings one
 obtains three other ONE diagrams depicted in figure 1({\it d-f}) in
 additional to the plane
 wave one (figure 1({\it a})). Assuming elastic rescattering to dominate,
 the pp-pair  is considered here
 as a quasibound system  with a fixed internal energy $E_{pp}$.
 Since the SS- and $\Delta$-mechanisms
 are less important they are treated  in the plane wave
 approximation.

 Numerical calculations are performed here with the RSC
 potential. As was shown in \cite{smuz98}, the result with the Paris
 potential obtained for the ONE-mechanism  is  very similar.
  The results for the cross section are shown in
 figure 3({\it a}). One can see from this figure that the
 $\Delta$ and SS- contributions have their maxima at $T_p\sim 0.6$ GeV,
 i.e. close to the dip of the ONE cross section. Nevertheless the coherent
 sum ONE+SS+$\Delta$ demonstrates well pronounced dip in the cross section.
  The ONE mechanism dominates at $T_p<0.5$ GeV and above 1 GeV.
 At $T_p>1 $GeV the ONE+$\Delta$+SS-model predicts a plateau
 in the cross section as a function of the initial energy
 at $\theta_{cm}=180^\circ$.
 This plateau manifests the $T_p$ dependence of the
 right hand side of Eq. (\ref{onecs}).

 The role of the $\Delta $-mechanism is  important mainly  in
 the node region $T_p\sim $ 0.7 GeV, but becomes  negligible at
 $T_p>1$ GeV. A  minor contribution is expected also at $T_p>1$ GeV
 from  excitation  of heavier nucleon isobars $N^*$   because of the
 same   suppressing isospin factors  as for the
 $\Delta$-isobar\footnote{An only exclusion might
 be the $N^*(1535)$-isobar strongly
 coupled to the $\eta-$meson. The maximum of the $N^*(1535)$-
 contribution to the reaction  (\ref{reaction1}) is
 expected at $T_p\sim 1.6 $ GeV.}.
 As seen from the $E_{pp}$ dependence of $t_s(q,k)$ (figure 2),
 the maximal value of the cross section of the reaction (\ref{reaction1})
 is expected at $E_{pp}=0.3-0.7$ MeV.
 At $E_{pp}<0.3$ MeV the cross section decreases rapidly
 due to Coulomb repulsion in the pp-system.
  At higher relative energies  $E_{pp}\sim 5-10$ MeV
 the role of nonzero orbital momenta $l\not=0$ in the half-off-shell
 NN-amplitude increases  and
 makes the node of the S-wave  amplitude $t_s(q,k)$  be non-visible
 in the cross section \cite{smuz98}.

  Rescatterings in the initial and final states fill in in part
  the ONE-minimum of the cross section.
  Nevertheless,  this minimum is well pronounced 
 (figure 3({\it a,c})).
  As  shown in figure 3({\it  b}),
  rescatterings produce a remarkable structure  in $T^{ONE}_{20}$
  in the region of the node of the half-off-shell NN-amplitude.
  On the contrary,  in the elastic $pd\to dp$  process
   the rescatterings do not change practically
  the tensor analyzing power $T^{ONE}_{20}(180^\circ)$ since in this case
  the node in  the $pn\to d$ vertex  of the
  ONE-diagram  is hidden by the D-wave contribution \cite{uz98}.
  Therefore  rescatterings in the reaction (\ref{reaction1}) and
  interference between the ONE and other mechanisms
  give an additional signal for the presence of the node.
   A similar irregular  behaviour demonstrates the
  spin-correlation parameter $C_{yy}$  depicted in figure 2({\it d})
  (for the definition see  \cite{ohlssen}). 

 The reaction (\ref{reaction1}) was not yet investigated experimentally.
  Available  experimental data  on the formation of the
 singlet $(pn)_s$ pair in $pd$ interactions
 are obtained in semiinclusive experiments  \cite {dubna,bfw}.
  The  energy resolution  in  \cite{dubna,bfw} was not
 enough high to observe  the above discussed properties of the 
 $p\,d\to (pn)_s\,p$ channel.
 The only exclusive measurement \cite{witten} of the reaction
 $pd\to pnp$  at 585 MeV displays a  very small
 fraction (about a few percent)  of the singlet contribution \cite{uzkfs}.
 Direct measurement of the singlet channel in the reaction 
 (\ref{reaction1}) is planned at COSY \cite{cosy}.

 In conclusion, we have found that the role
 of the  ONE mechanism increases  considerably
 in  the reaction (\ref{reaction1}) in comparison with  
 the process $pd\to dp$. It caused  i) by
 the isotopic spin factors
 suppressing  the  diagrams which are relevant 
 to excitation of the nucleon isobars in the intermediate
 state of this reaction and ii) by  dominance of the S-state
 in the pp-system at low $E_{pp}$. As a result, we found within
 the DWBA ONE + $\Delta$ + SS model that
 the node available in  the standard potential picture of
 the half-off-shell pp($^1S_0)$ scattering
 amplitude
 appears as an  irregularity in the behaviour of the observables.
 Thus, the reaction (\ref{reaction1})
 provides   new qualitative criteria to clarify  the role 
 of nucleon degrees of freedom in NN-system at short distances.

 {\bf Acknowledgments}. The author would like to thank V.I. Komarov,
 F. Rathmann, H. Seyfarth and H. Str\"oher for helpful remarks 
 and   warm hospitality at IKP of Forschungszentrum J\"ulich  were 
 a part of this work was done. 


\eject





\eject
\input epsf
\begin{figure}[t]
\begin{center}
\vspace*{-1cm}
\mbox{\epsfxsize=6.0in \epsfbox{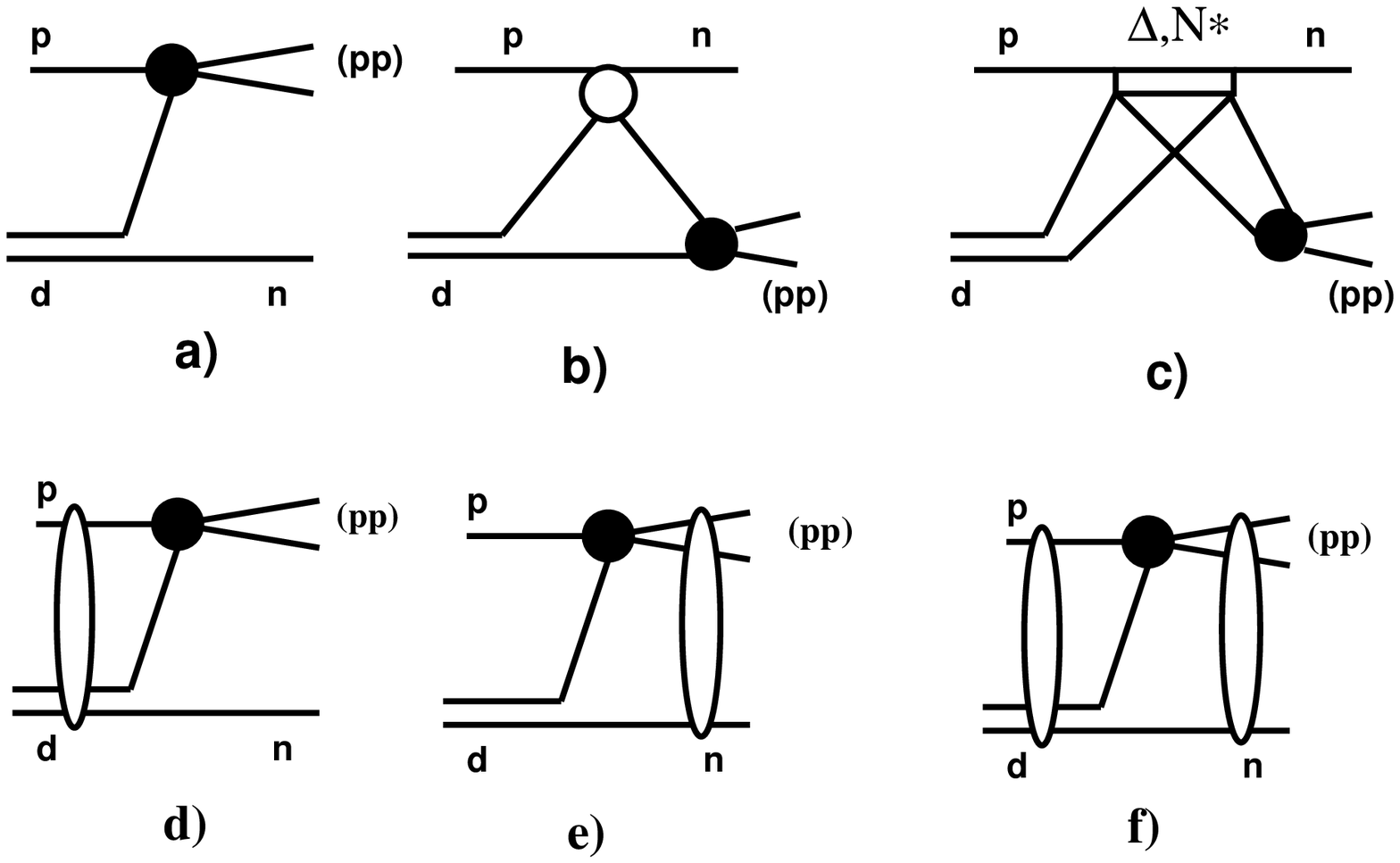}}
\caption {Mechanisms of the reaction $p+d\protect\to n+pp$:
       $ (a) $  -- one-nucleon exchange (ONE),
       $ (b)$ -- single scattering (SS),
       $ (c)$ -- double pN-scattering $(\Delta )$  with excitation
     of the $\Delta-$  or $N^*-$ isobar.
     The rescatterings are shown  for the one-nucleon exchange 
     in the initial $(d)$, final $(e)$ and initial plus final $(f)$
     states.
 }
\label{uzi-1}
\end{center}
\end{figure}

\eject
\begin{figure}
\begin{center}
\vspace*{-1cm}
\mbox{\epsfxsize=6.0in \epsfbox{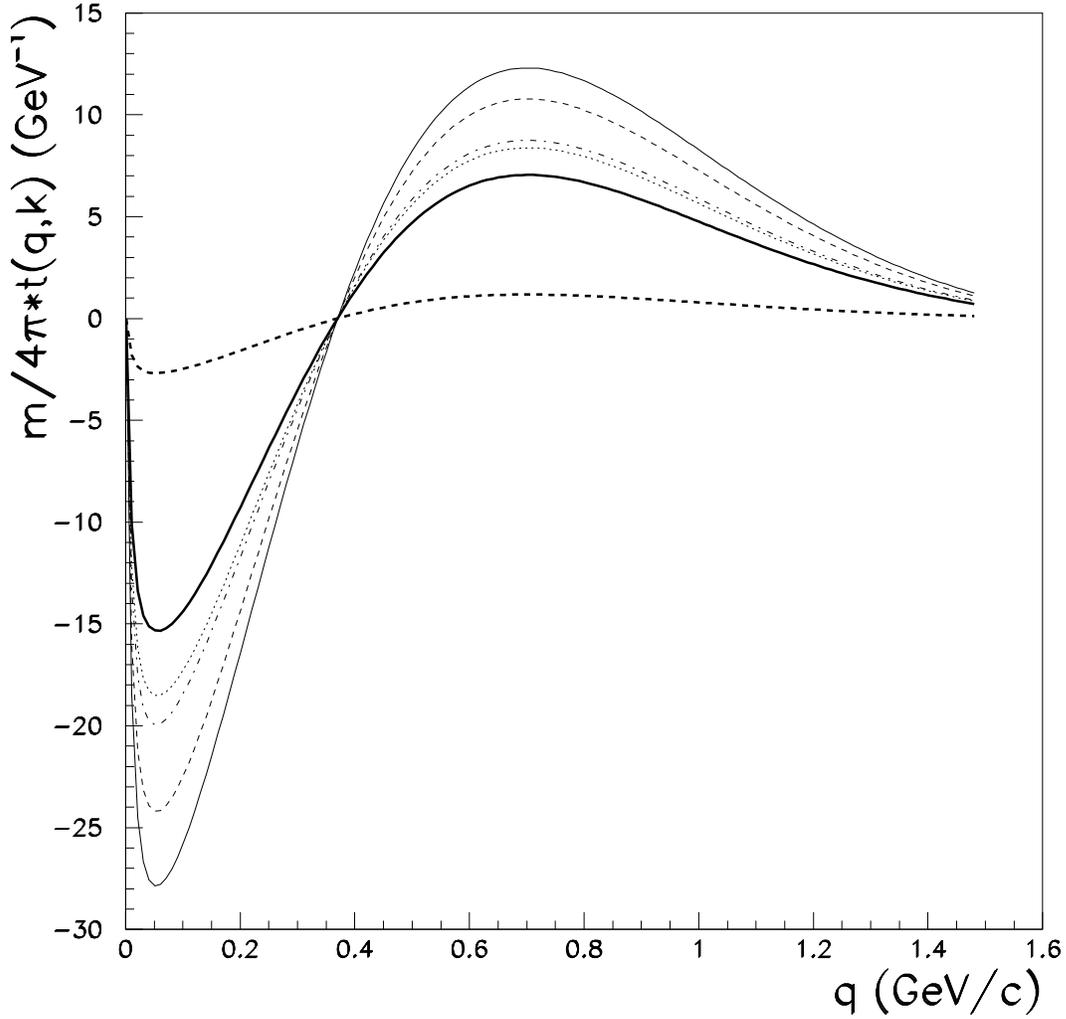}}
\caption
{The half-off-shell $pp(^1S_0)$ scattering amplitude $({m}/{4\pi})\,t(q,k)$
  as a function of the off-shell momentum $q$ at different energies
 $E_{pp}$: 0.01 MeV (dashed thick line), 0.1 MeV 
 (dashed-dotted), 0.5 MeV (full thin), 1.0 MeV (dashed), 2.0 MeV (dotted),
 3 MeV (full thick). The on-shell amplitude  $t(k,k)$ is related to the 
 Coulomb-nuclear phase shift $\delta$ in the $^1S_0$ state  as
$({m}/{4\pi})\,t(k,k)=-\frac{1}{k}\exp{[i\delta]}\sin{\delta}$. }
\label{uz2}
\end{center}
\end{figure}
\eject

\begin{figure}
\begin{center}
\vspace*{-1cm}
\mbox{\epsfxsize=6.0in \epsfbox{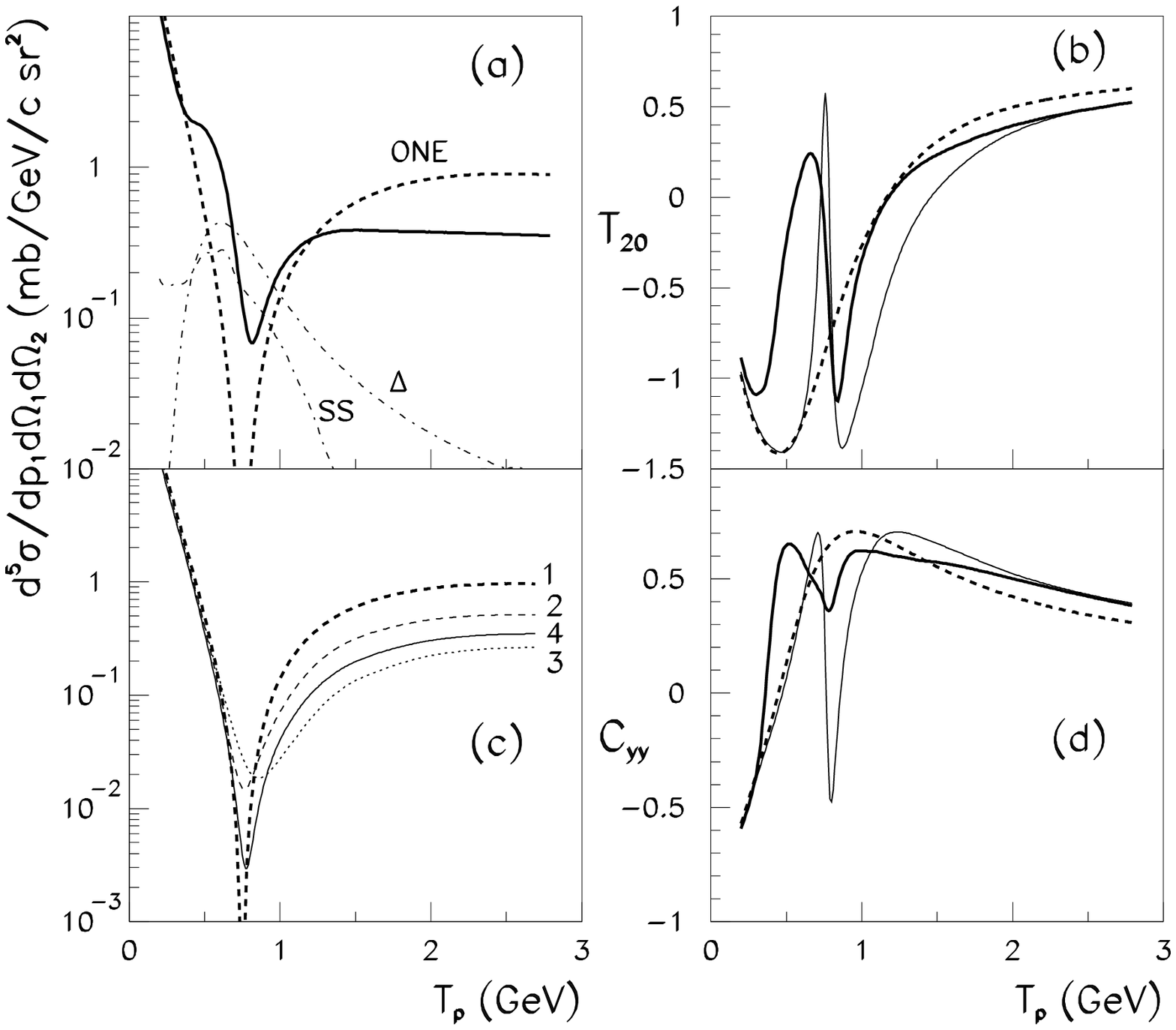}}
\caption 
{ The  laboratory cross section ({\it a,c}),  tensor analyzing power
 $T_{20}$ ({\it b}) and spin-spin correlation parameter $C_{yy}$ ({\it d})
 of the  reaction (\protect\ref{reaction1})
 versus the kinetic energy of the proton
 beam  $T_p$ at the neutron scattering angle
 $\theta_{cm}=180^\circ$ and the internal energy of the $pp$ pair
 $E_{pp}=3$ MeV for the different mechanisms: ONE without
 rescatterings (dashed thick line), ONE  with all  rescatterings (DWBA ONE)
 in the initial and final states  (full thin), the coherent sum of
 the DWBA ONE+  $\Delta$ +SS mechanisms (full thick). 
 The  SS and $\Delta$ contributions   are shown on the panel {\it a}
 by dash-dotted curves. 
 In the panel  {\it c}, the curves  correspond to 
 the  ONE diagrams   depicted in figure 1: 1 -- $a$,
 2 -- $a+d$, 3 -- $a+d+e$,
 4 -- $a+d+e+f$ (DWBA ONE).
}
\label{secit20}
\end{center}
\end{figure}

\begin{thebibliography}{99}
\bibitem{dubnainclus}
 Azhgirey L S {\it et al} 1996 {\it Phys. Lett.} B {\bf 387} 37
\bibitem{dubna}  Azhgirey L S {\it et al} 1998 {\it Yad. Fiz.} {\bf 61} 494
\bibitem{perdrisat90}  Perdrisat C F and   Punjabi V 1990 {\it  Phys. Rev.}
 C {\bf 42} 1899
\bibitem{abott} Abott D et al. 2000 {\it  Phys. Rev. Lett.} {\bf 84}  5053
\bibitem{rsc}  Reid J R V 1968 {\it Ann. Phys. (N.Y.)} {\bf 50} 411
\bibitem{paris}  Lacombe M {\it et al} 1981 {\it Phys. Lett.} {\bf B101} 139
\bibitem{bonn}  Machleidt R {\it et al} 1987 {\it Phys. Rep.} {\bf 149}  1
\bibitem{kk}  Kerman A K and  Kisslindger L S 1969 {\it Phys. Rev.} {\bf 180}
 1483
\bibitem{kob98} Kobushkin A P 1998 {\it Phys. Lett.} B {\bf 421} 53 
\bibitem{lak}  Kondratyuk L A, Lev F M and  Shevchenko L V 1981
 {\it Yad. Fiz.} {\bf 33} 1208
\bibitem{aleshin}  Aleshin N P {\it et al} 1994 {\it Nucl. Phys.}
 A {\bf 568} 809;
  Belostotski S L {\it et al} 1997 {\it Phys. Rev.} C {\bf 56} 50
\bibitem{kolybas}  Craigie N S  and  Wilkin C  1969 {\it  Nucl. Phys.}
 B {\bf 14} 477;
  Kolybasov V M and  Smorodinskaya N Ya 1973 {\it Yad. Fiz. } {\bf 17} 1211;
  Nakamura A and  Satta L 1985 {\it  Nucl. Phys.} A { \bf 445} 706
\bibitem{bdillig} Boudard A and Dillig M 1985 {\it Phys. Rev.} C {\bf 31}
 302
\bibitem{uz98}  Uzikov Yu N 1998 {\it Phys. Part. Nucl.} {\bf 29} 583
\bibitem{uz2000}  Uzikov Yu N 2000 JINR Preprint E2-200-149,
 nucl-th/0006067; 2001  JINR Preprint E4-2001-237, nucl-th/0111079 
\bibitem{smuz98}  Smirnov A V and  Uzikov Yu N 1998 {\it Phys. At. Nucl.}
 {\bf 61} 361
\bibitem{imuz90} Imambekov O and  Uzikov Yu N 1990
  {\it Sov. J. Nucl. Phys.} {\bf 52} 862
\bibitem{imuz88} Imambekov O and  Uzikov Yu N 1988 {\it  Yad. Fiz.}
 {\bf 47} 1089
\bibitem{habitzsch} Hudomaly-Gabitzsch J {\it et al} 1978 {\it Phys. Rev.}
 {\bf C18} 2666
\bibitem{bfw} Boudard A, F\"aldt G and  Wilkin C 1996 {\it Phys. Lett.}
 B {\bf 389} 440
\bibitem{blu} Blokhintsev L D, Lado A V and  Uzikov Yu N 1996
 {\it Nucl. Phys.} A{\bf 597} 487
\bibitem{uzpddp98}  Uzikov Yu N 1998 {\it Nucl. Phys.} A {\bf 644} 321
\bibitem{nikhef} Bouwhuis M {\it et al} 1999
 {\it Phys. Rev. Lett.} {\bf 82} 3755
\bibitem{ohlssen} Ohlsen G 1972 {\it Rep. Prog. Phys.} {\bf 35} 717
\bibitem{witten} Witten T {\it et al} 1975  {\it Nucl. Phys.} A {\bf 254}
 269
\bibitem{uzkfs}  Uzikov Yu N {\it et al}  nucl-th/0110038, 
 {\it Phys. Lett.} B (in press)
\bibitem{cosy} Beam time request to COSY proposal $N^o$ 20 (1999)
 Spokesperson Komarov 
~V~I,
 www.ikpd15.ikp.fz-juelich.de:8085/doc/Publications.html


\end{thebibliography}
\end{document}